\documentclass[aps,pre,showpacs,preprint]{revtex4}

\usepackage{amsmath}
\usepackage{amssymb}
\usepackage[utf8]{inputenc}
\usepackage{graphicx}
\usepackage{dcolumn}
\usepackage[mathscr]{euscript}
\usepackage{bm}
 
\DeclareMathOperator\erfi{erfi}

\begin{document}

\title{  Self-diffusion in a quasi-two dimensional gas of hard spheres }
\author{J. Javier Brey, M. I. Garc\'{\i}a de Soria, and P. Maynar}
\affiliation{F\'{\i}sica Te\'{o}rica, Universidad de Sevilla,
Apartado de Correos 1065, E-41080, Sevilla, Spain}
\date{\today }

\begin{abstract}
A quasi-two-dimensional system of hard spheres strongly confined between two parallel plates is considered. The attention is focussed on the macroscopic self-diffusion process observed when the system is looked from above or from below. The transport equation, and the associated self-diffusion coefficient, are derived from a Boltzmann-Lorentz kinetic equation, valid in the dilute limit.  Since the equilibrium state of the system is inhomogeneous, this requires the use of a modified Chapman-Enskog expansion that distinguishes between equilibrium and non-equilibrium gradients of the density of labelled particles. The  self-diffusion coefficient is obtained as a function of the separation between the two confining plates. The theoretical predictions are compared with molecular dynamics simulation results and a good agreement is found.

\end{abstract}

\maketitle

\section{Introduction}
\label{s1}
The interest on gas and liquid microflows has largely increased in the last decades. This has been motivated by the development of new technologies allowing the construction and manipulation of microfluidics devices. The efficiency of these devices depends rather strongly on the properties of the transport processes taking place in the fluid. Self-diffusion is one of these fundamental processes.  For instance, the analysis of self-diffusion provides relevant information on the structure of porous media. Nevertheless, its experimental measurement, and also of the other transport coefficients,  in  small channels or porous presents many difficulties.  Much of what is known follows from molecular dynamics simulation results \cite{AyR11,RBEyI11,Metal08}, although there have been also proposals of a general statistical mechanics theory for transport processes of fluids under confined conditions \cite{RyB15}. 

In this paper, self-diffusion is studied in a strongly confined dilute gas of gas spheres, namely the space accesible to the particles is limited by two parallel  plates separated a distance smaller than two particle diameters.  As a consequence of this geometry, particles cannot  jump on to each other and the system can be considered as quasi-two-dimensional (Q2D). Although it is known  that in real devices the nature of the walls plays a crucial role in determining the self-diffusion process \cite{RyB15}, in this study the interest focusses in isolating the effect of the confinement, as a restriction of the space accesible to the system. For this reason, the simplest model of confining walls is considered: elastic hard walls. For this system, a Boltzmann kinetic equation has been derived \cite{BMyG16,BGyM17}. The equation obeys a modified $H$-theorem implying the approach to the equilibrium state from any arbitrary initial condition.  Although the equilibrium velocity distribution is Gaussian, the density is not uniform, as a consequence of the confinement. There is a density gradient in the direction perpendicular to the confining walls. The equilibrium predictions following from the Boltzmann equation agree with results derived by means of equilibrium statistical mechanics \cite{SyL96,SyL97} and also with molecular dynamics simulation measurements \cite{BMyG16}. 

From the Boltzmann equation,  is is easy to construct the corresponding Boltzmann-Lorentz equation describing self-diffusion at equilibrium. Once a kinetic equation has been formulated to describe the dynamics of a defined system, there is a well established procedure to derive macroscopic or hydrodynamic transport equations from it \cite{RydL77}.  Although the method has been mainly applied to study transport processes in the bulk of a system, and those processes can be quite different in highly confined geometries, the expectation is that the method  can be adapted to describe also macroscopic transport in confined systems. On the other hand, one must be aware that there are significant differences to confront. The equilibrium state of a strongly confined system is not homogeneous, as mentioned, so that the presence of gradients can not be identified always with the existence of non-equilibrium macroscopic flows.  Moreover, hydrodynamic length scales, and hence hydrodynamic behavior, is not to be expected in the directions of high confinement, in our particular system, perpendicular to the confining plates. Hydrodynamic-like equations can be useful only to describe the macroscopic motion of the fluid in the directions in which particles of the fluid  can move distances much larger that the mean free path.

Self-diffusion is considered as the prototype of transport processes. For bulk systems of hard spheres, it has been the testing ground for many proposed approximations introduced in the context of non-equilibrium statistical mechanics. It is worth to mention that the methods have also been extended to systems composed of inelastic hard spheres \cite{BRCyG00,DByL02,LByD02}, although in this case there is no equilibrium state and the homogeneous reference state in which self-diffusion is studied is time-dependent. This has some similarity with confined systems, as the one considered here, in which the equilibrium reference state is inhomogeneous.

In the next section, the Boltzmann equation for the confined Q2D is reviewed, and particularized for self-diffusion processes in equilibrium by differentiating  between labeled  and unlabeled particles. The expression for the equilibrium density profile along the direction perpendicular to the plates is indicated. From the kinetic equation, the exact balance equation for the number density of labeled particles is derived. The equation involves the flux of particles, that in order to get a closed description, must be expressed in terms of the corresponding density field. The formal procedure to do the above is discussed in Sec.\ \ref{s3}, by means of a modified Chapman-Enskog expansion, that takes into account the density inhomogeneity of the equilibrium state.  In its original formulation \cite{CyC95,McL89}, the idea of the Chapman-Enskog method is to obtain a ``normal solution'' of the kinetic equation in the form of a gradient expansion around a local-equilibrium state. A normal solution has the property that all its space and time dependence occurs through the macroscopic hydrodynamic fields. In the case of self-diffusion, the only hydrodynamic field is the number density of labeled particles. Since, in the particular system we are considering, the stationary equilibrium state is not homogeneous, it is clear that the existence of a density gradient is not enough to imply the presence of a macroscopic flow of particles in the system. In other words, the zeroth order approximation in the expansion must incorporate those gradients that are associated to equilibrium inhomogeneities.  Then,  the concept of normal solution itself must be generalized. The general issue of formulating a modified Chapman-Enskog expansion for fluctuations about a non-equilibrium state has been addressed in a seminal paper by Lutsko \cite{Lu06}.  The expansion employed here differs for the method developed in that work. A short comparison of both approaches is presented in appendix \ref{ap1}. The reason for the difference is that we are dealing with an equilibrium state and not with a non-equilibrium situation. On the other hand, 
it must be stressed that the study of transport processes in systems exhibiting an inhomogeneous steady state also requires a modification of the usual Champan- Enskog expansion, as it is discussed in this paper.

The practical application of the modified Chapman-Enskog method requires, as the original one, to make some kind of approximation to get and explicit expression for the self-diffusion coefficient.  Usually, an expansion in some complete set of orthogonal polynomials is used, and only the leading term is kept often. In addition, to render  the mathematical complexity following from the confinement analytically tractable, an expansion in the separation of the two plates is performed. Both approximations are discussed in Sec.\ \ref{s4}, where the resulting expression for the self-diffusion coefficient is given. The theoretical predictions are compared with molecular dynamics simulation results in Sec.\ \ref{s5}. There, it is shown that the projection of the motion of the particles on a plane parallel to the plates has a diffusive character. 
The diffusion coefficient is measured from the simulation data for the mean square displacement of the particles as a function of time. The obtained values are in good agreement with the theoretical expression. The paper ends in Sec. \ref{s6} with a short summary of the results and a discussion of possible extensions. Appendices \ref{ap2} and \ref{ap3}, contain details of the calculations mentioned along the paper.

\section{The kinetic equation and  the conservation law}
\label{s2}
The system considered is composed of $N$ identical hard spheres of diameter $\sigma$ and mass $m$, confined between two parallel hard plates located at $z=0$ and $z=h$, respectively, being $\sigma \leq  h  < 2 \sigma$. No external force is acting on the particles. In particular, the effect of gravity is assumed to be negligible. Collisions between particles and also of the particles with the plates are elastic. Although all the spheres are mechanically identical, $N_{l}$ of them have a label, or color, that differentiate them from the others. The one-particle distribution function of labelled particles, providing the density of particles at a given position ${\bm r}$ and with a given velocity ${\bm v}$ at time $t$, will be denoted by $f_{l}({\bm r},{\bm v},t)$, while the one-particle distribution function for all particles, labelled and unlabelled, will be $f({\bm r},{\bm v},t)$. Here attention will be restricted to situations in which the system is very dilute and at equilibrium with a temperature $T$. Then, it is known that \cite{SyL96,SyL97}
\begin{equation}
\label{2.1}
f({\bm r},{\bm v},t) \rightarrow f_{eq} (z,{\bm v})= n_{0}(z) \varphi_{MB}({\bm v}),
\end{equation}
where $\varphi_{MB}$ is the Maxwellian distribution,
\begin{equation}
\label{2.2}
\varphi_{MB}({\bm v}) =  \left( \frac{m}{2 \pi k_{B} T} \right)^{3/2} e^{-\frac{mv^{2}}{2k_{B}T}}
\end{equation}
and
\begin{equation}
\label{2.3} 
n_{0}(z) = \frac{N}{Ab} \exp \left[ a \left( z-\frac{h}{2} \right) ^{2} \right]
\end{equation}
with
\begin{equation}
\label{2.4}
b = \left( \frac{\pi}{a} \right)^{1/2} \erfi \left[ \frac{\sqrt{a} (h-\sigma)}{2} \right].
\end{equation}
In the above expressions, $k_{B}$ is the Boltzmann constant, $A$ is the area of each of the parallel plates, $a=\pi N/A$, and $\erfi(x)$ is the imaginary error function,
\begin{equation}
\label{2.5}
\erfi (x) \equiv \pi^{-1/2} \int_{-x}^{x} dy\, e^{y^{2}}.
\end{equation}
The density profile given by Eq.\, (\ref{2.3}) is a consequence of the confinement of the system and it has been derived by means of both equilibrium statistical mechanics methods \cite{SyL96} and kinetic theory \cite{BMyG16,BGyM17}. Moreover, it has been shown to be in good agreement with molecular dynamics simulation results \cite{BMyG16}.

The distribution function of labeled particles obeys a modified Boltzmann-Lorentz (BL) equation that follows by reproducing step by step the arguments used in Ref. \cite{BGyM17}. Taking into account that labeled particles collide with  labeled as well as unlabeled particles and that the total system is at equilibrium, it is readily obtained
\begin{equation}
\label{2.6}
\left( \frac{\partial}{\partial t}  + {\bm v} \cdot {\bm \nabla}  \right) f_{l}({\bm r},{\bm v}, t)=  \Lambda \left[ {\bm r},{\bm v} |f_{eq}\right] f_{l}({\bm r}, {\bm v},t),
\end{equation}
where $\Lambda$ is the modified BL linear collision operator defined by 
\begin{eqnarray}
\label{2.7}
\Lambda \left[ {\bm r},{\bm v} |f_{eq}\right] \phi ({\bm r},{\bm v})  &\equiv & \sigma \int d{\bm v}_{1} \int_{\sigma/2}^{h-\sigma/2} dz_{1}\,  \int_{0}^{2\pi} d \psi\, |  {\bm g} \cdot \widehat{\bm \sigma} | \ \nonumber \\
&& \times  \left[ \Theta ({\bm g} \cdot \widehat{\bm \sigma}) f_{eq} (z_{1}, {\bm v}^{\prime}_{1}) \phi ({\bm r},{\bm v}^{\prime}) -  \Theta (-{\bm g} \cdot  \widehat{\bm \sigma}) f_{eq} (z_{1}, {\bm v}_{1}) \phi ({\bm r},{\bm v}) \right],
\end{eqnarray}
for arbitrary $\phi({\bm r}, {\bm v})$.  Here ${\bm g} \equiv {\bm v}_{1}-{\bm v}$ is the relative velocity of the two colliding particles prior to  the collision, $\Theta$ is the Heaviside step function, ${\bm v}^{\prime}$ and ${\bm v}_{1}^{\prime}$ are the velocities of the two particles after the collision defined by the unit vector $\widehat{\bm \sigma}$ joining the centers of the two particles at contact, so that
\begin{eqnarray}
\label{2.8}
{\bm v}^{\prime} &=& {\bm v} + ({\bm g} \cdot \widehat {\bm \sigma}) \widehat{\bm \sigma}  \nonumber \\
{\bm v}^{\prime}_{1} &=& {\bm v}_{1} - ({\bm g} \cdot \widehat {\bm \sigma}) \widehat{\bm \sigma} .
\end{eqnarray}
In the coordinate system we are using, it is
\begin{equation}
\label{2.9}
\widehat{\bm \sigma} = \left\{ \sin \theta \sin \psi, \sin \theta \cos  \psi, \cos \theta \right\}
\end{equation}
with
\begin{equation}
\label{2.10}
\cos \theta = \frac{z_{1}-z }{\sigma}\, , \quad \sin \theta \geq 0 .
\end{equation}
The operator $\Lambda$ has the relevant property that, for any pair of functions $\phi ({\bm v})$ and $\chi ({\bm v})$, it is
\begin{eqnarray}
\label{2.11}
\int d{\bm v}\,  \phi ({\bm v}) \Lambda[{\bm r},{\bm v}| f_{eq}] \chi ({\bm v})  &= & \sigma \int d{\bm v} \int d{\bm v}_{1} \int_{\sigma/2}^{h-\sigma/2} dz_{1}   \int_{0}^{2\pi} d \psi\, |  {\bm g} \cdot \widehat{\bm \sigma}| \Theta (-{\bm g} \cdot \widehat{\bm \sigma}) f_{eq}(z,{\bm v}_{1}) \nonumber \\
&& \times  \chi ({\bm v})   \left[ \phi ({\bm v}^{\prime}) -\phi ({\bm v}) \right]
\end{eqnarray}
Since $f_{eq} (z,{\bm v})$ is the steady solution of the nonlinear Boltzmann equation for the confined gas, it verifies \cite{BMyG16}
\begin{equation}
\label{2.12}
v_{z} \frac{\partial}{\partial z}\, f_{eq}(z,{\bm v}) = \Lambda[{\bm r},{\bm v}|f_{eq}] f_{eq}(z,{\bm v}).
\end{equation}
Equation (\ref{2.6}) holds inside the system and it must be complemented by appropriate boundary conditions. In our case, those conditions must be consistent with the form of the equilibrium distribution function given in Eq. (\ref{2.1}). A more detailed discussion of this issue is given in ref. \cite{BGyM17}.
Labeled particles interchange momentum and kinetic energy with the rest of particles in the system and, therefore, these quantities are not collisional invariants of the BL operator $\Lambda$. Only the number of labeled particles is conserved and, consequently, its time evolution obeys a conservation law. Define the number density of labeled particles, $n_{l}({\bm r},t)$  in the usual way
\begin{equation}
\label{2.13}
n_{l}({\bm r}, t) \equiv \int d{\bm v}\, f_{l}({\bm r}, {\bm v},t).
\end{equation}
Velocity integration of the BL equation leads to the conservation law
\begin{equation}
\label{2.14}
\frac{\partial}{\partial t}\, n_{l}({\bm r},t) + {\bm \nabla} \cdot {\bm J}_{l} ({\bm r}, t) =0,
\end{equation}
where the flux of labeled particles, ${\bm J}_{l}$, is defined by
\begin{equation}
\label{2.15}
{\bm J}_{l}({\bm r},t)= \int d{\bm v}\, {\bm v} f_{l}({\bm r},{\bm v},t).
\end{equation}
Upon deriving Eq. (\ref{2.13}), the property given in Eq. (\ref{2.11}) has been used. Of course, the BL equation  admits the steady solution
\begin{equation}
\label{2.16}
f_{l,eq}(z,{\bm v}) = n_{l,0}(z) \varphi_{MB} ({\bm v}),
\end{equation}
with
\begin{equation}
\label{2.17}
n_{l,0}(z) = \frac{N_{l}}{N}\, n_{0}(z) = \frac{N_{l}}{Ab} \exp \left[ a \left(z-\frac{h}{2} \right)^{2} \right],
\end{equation}
describing the long time equilibrium state, in which the labeled particles are  spatially distributed in the same way as the unlabeled ones.. 
The aim here is to derive an equation for the transport of labeled particles occurring in the horizontal plane, i.e. parallel to the confining plates. In other words, we are interested in the macroscopic motion of  labeled particles when seen from above or from bellow. Then, we define the number density of labelled particles in the horizontal plane by
\begin{equation}
\label{2.18}
n_{l=}({\bm r}_{=},t)= \int_{\sigma/2}^{h-\sigma/2} dz\,  n_{l}({\bm r},t),
\end{equation}
where ${\bm r}_{=} $ denotes the vector projection of ${\bm r}$  on a plane parallel to the plates. From Eq. (\ref{2.14}), it follows that
\begin{equation}
\label{2.19}
\frac{\partial}{\partial t}\, n_{l=}({\bm r}_{=},t) =- \int_{\sigma/2}^{h-\sigma/2} dz\, {\bm \nabla} \cdot  {\bm J}_{l} ({\bm r}, t)
\end{equation}
and, taking into account that $J_{z}$ vanishes at $z=\sigma/2$ and $z=h-\sigma/2$ due to the hard walls \cite{MGyB18,BMyG19a}, the above equation becomes
\begin{equation}
\label{2.20}
\frac{\partial}{\partial t} n_{l=} ({\bm r}_{=},t) =- {\bm \nabla}_{=} \cdot  \boldsymbol{\mathcal{J}}_{l=} ({\bm r}_{=},t),
\end{equation}
with
\begin{equation}
\label{2.21}
 \boldsymbol{\mathcal{J}}_{l=} ({\bm r}_{=},t) = \int_{\sigma/2}^{h-\sigma/2} dz  \int d{\bm v}\, {\bm v}_{=} f_{l}({\bm r},{\bm v},t)
 \end{equation}
 and ${\bm \nabla}_{=}$ defined in the plane $z={\text const.}$ To convert Eq.\, (\ref{2.20}) into a closed equation for the horizontal density of labeled particles, one must derive an expression for  the flux $\boldsymbol{\mathcal J}_{=}$ in terms of $n_{l=}({\bm r}_{=},t)$ and its spatial derivatives. The standard procedure to do so starting from a kinetic equation, is the Chapman-Enskog expansion \cite{CyC95,McL89}, whose goal is to construct a so-called normal solution of the kinetic equation. The latter is defined as a distribution function in which all the space and time dependence occurs through the hydrodynamic fields associated to the conserved quantities. In the present case, the only hydrodynamic field is the number density $n_{l}({\bm r},t)$ and, therefore, a normal distribution has the form
 \begin{equation}
 \label{2.22}
 f_{l}({\bm r},{\bm v},t)= f_{l}[{\bm v}|n_{l}({\bm r},t)].
 \end{equation}
This is a functional dependence, so that gradients of all orders of $n_{l}$ are involved. Notice that the normal distribution associated to the kinetic equation (\ref{2.5}) is a functional of $n_{l}$, while in order to close Eq. (\ref{2.19}) we need to express $\boldsymbol{\mathcal J}_{=}$ in terms of $n_{l=}$. It will be later seen how this occurs due to  the peculiarities of the geometry of the system we are dealing with.

The normal form of the distribution function is assumed to be reached from any arbitrary initial distribution, $f_{l}({\bm r},{\bm v},0)$, for large enough times. To find the normal solution, the Chapman-Enskog algorithm uses  perturbation expansion in the gradients of the hydrodynamic fields. Transport phenomena occur due to deviations from the equilibrium state, that in the present case happens to be inhomogeneous.  Therefore, gradient expansion of $n_{l,0}(z)$ will be avoided. On the other hand, it is important to establish that the method to be used allows for arbitrary large deviations of $n_{l}({\bm r},t)$ from its equilibrium value $n_{l,0}(z)$. In particular, no linearization around $\delta n_{l} \equiv n_{l}-n_{l,0}$ will be made.

\section{The modified Chapman-Enskog expansion around the inhomogeneous equilibrium state}
\label{s3}
Let us introduce a dimensionless function, $\nu({\bm r},t)$, defined by
\begin{equation}
\label{3.1}
n_{l}({\bm r},t)= n_{l,0}(z) \nu ({\bm r},t),
\end{equation}
so that at equilibrium it is $\nu=1$. To formulate the perturbation theory we are going to develop, it is convenient to introduce a formal uniformity parameter $\epsilon$, and to decompose the action of the gradient operator on the labeled particles  density field as
\begin{equation}
\label{3.2}
{\bm \nabla} n_{l}({\bm r},t) = {\bm \nabla}^{(0)} n_{l}({\bm r},t) + \epsilon {\bm \nabla}^{(1)} n_{l}({\bm r},t),
\end{equation}
where, by definition, it is
\begin{equation}
\label{3.3}
{\bm \nabla}^{(0)} n_{l}({\bm r},t) \equiv  \nu ({\bm r},t) {\bm \nabla} n_{l,0}(z) = \nu ({\bm r},t) \frac{dn_{l,0}(z)}{dz}\, \widehat{\bm e}_{z},
\end{equation}
\begin{equation}
\label{3.4}
 {\bm \nabla}^{(1)} n_{l}({\bm r},t)  \equiv n_{l,0}(z) {\bm \nabla} \nu ({\bm r},t).
 \end{equation}
In Eq.\ (\ref{3.3}), the unit vector along the positive direction of the $z$-axis, $\widehat{\bm e}_{z}$, has been introduced.

Given the form of the normal distribution we are looking for, the above separation of the gradient operator will generate an expansion of the one-particle distribution function looking like 
\begin{equation}
\label{3.5}
f_{l}[{\bm v}|n_{l}({\bm r},t)]=f_{l}^{(0)}[{\bm v}|n_{l}({\bm r},t)] + \epsilon f_{l}^{(1)}[{\bm v}|n_{l}({\bm r},t)]+ \epsilon^{2} f_{l}^{(2)}[{\bm v}|n_{l}({\bm r},t)] + \ldots,
\end{equation}
where $f_{l}^{(0)}$ is of zeroth order in ${\bm \nabla} \nu$, $f_{l}^{(1)}$ is linear in ${\bm \nabla} \nu$, $f_{l}^{(2)}$ is linear in  $\nabla^{2} \nu$ and $({\bm \nabla} \nu)^{2}$, etc. On the other hand, at each order of the perturbation, the distribution can be a function of the exact density field, $n_{l}({\bm r},t)$, as well as all gradients of the equilibrium density, $n_{l,0}(z)$.

The expansion of the one-particle distribution function generates a similar one for the flux of labeled particles,
\begin{equation}
\label{3.6}
{\bm J}_{l}({\bm r},t)= \sum_{j=0}^{\infty} \epsilon^{j} {\bm J}_{l}^{(j)} ({\bm r},t),
\end{equation}
\begin{equation}
\label{3.7} {\bm J}_{l}^{(j)}({\bm r},t) \equiv \int d{\bm v}\, {\bm v} f_{l}^{(j)}({\bm r},{\bm v},t).
\end{equation}
Space and time derivatives of the distribution function are related by Eq.\ (\ref{2.14}) and, therefore it is necessary to carry out a multiscale expansion of the balance equation, and to write
\begin{equation}
\label{3.8}
\frac{\partial f_{l}}{\partial t} = \partial_{t}^{(0)} f_{l}+ \epsilon \partial_{t}^{(1)} f_{l} + \epsilon^{2} \partial_{t}^{(2)} f_{l} + \cdots.
\end{equation}
In this expansion, it is understood that the normal form of the distribution function being constructed and the balance equation (\ref{2.14}) are used to express the time derivative at each order as a function of the gradients of the density field. 

The zeroth order distribution, $f_{l}^{(0)}$, is defined such that it gives the exact value of the density field of labeled particles, i.e.
\begin{equation}
\label{3.9}
\int d{\bm v}\, f_{l}^{(0)} [{\bm v} |n_{l}({\bm r},t)] = n_{l}({\bm r},t)
\end{equation}
and, consistently, it must be
\begin{equation}
\label{3.10}
\int d{\bm v}\, f_{l}^{(j)} [{\bm v} |n_{l}({\bm r},t)] =0,
\end{equation}
for $j>0$, so that the definition in Eq.\ (\ref{2.13}) is preserved. 

Using the $\epsilon$-expansion of the several quantities generated above, it follows that the zeroth order kinetic equation is
\begin{equation}
\label{3.11}
\left(\partial_{t}^{(0)} f_{l}^{(0)}+{\bm v} \cdot {\bm \nabla}^{(0)}\right) f_{l}^{(0)} [{\bm v}|n({\bm r},t)]= \Lambda[{\bm r},{\bm v}|f_{eq} ]  f_{l}^{(0)} [{\bm v}|n({\bm r},t)]
\end{equation}
and the zeroth order balance equation is
\begin{equation}
\label{3.12}
\partial_{t}^{(0)} n_{l}+{\bm \nabla}^{(0)} \cdot {\bm J}_{s}^{(0)} =0.
\end{equation}
Taking into account the definition of the operator ${\bm \nabla}^{(0)}$ given in Eq. (\ref{3.3}), it is seen that Eq. (\ref{2.12}) implies
\begin{equation}
\label{3.13}
{\bm v} \cdot {\bm \nabla}^{(0)} \nu({\bm r},t) f_{l,eq}(z,{\bm v}) = \Lambda [{\bm r}, {\bm v}|f_{eq}]  \nu({\bm r},t) f_{l,eq}(z,{\bm v}),
\end{equation}
since the factor $\nu({\bm r},t)$ cncels out at both sides of the equation. Comparison of this equation with Eq.\ (\ref{3.11}) leds to identify
\begin{equation}
\label{3.14}
f_{l}^{(0)} [{\bm v}|n({\bm r},t)] = \nu ({\bm r},t) f_{l,eq}(z,{\bm v})= n_{l}({\bm r},t) \varphi_{MB}({\bm v}).
\end{equation}
as the normal solution of Eq.\ (\ref{3.11}). Indeed, with this identification it is
\begin{equation}
\label{3.15}
{\bm J}_{s}^{(0)}= \int d{\bm v}\, {\bm v} f_{l}^{(o)} [{\bm v}|n({\bm r},t)] =0
\end{equation}
and, hence from Eq. (\ref{3.12})
\begin{equation}
\label{3.16}
\partial_{t}^{(0)} n_{l}({\bm r},t)=0
\end{equation}
and, consequently,
\begin{equation}
\label{3.17}
\partial_{t}^{(0)} f_{l}^{(0)} [{\bm v}|n({\bm r},t)]=0.
\end{equation}
Moreover, the condition given by Eq. (\ref{3.9}) is trivially verified. 

The result in Eq.\ (\ref{3.14}) shows that the zeroth order distribution function is obtained from the equilibrium distribution of labeled particles  by replacing the equilibrium density profile, $n_{l,0}(z)$, by the actual non-equilibrium density field, $n_{l}({\bm r},t)$. This is the extension of the usual concept of local equilibrium to the present case. The simplicity of this result is due to the fact that we are considering an  equilibrium reference state, although it is inhomogeneous. The general issue of an expansion around an arbitrary non-equilibrium state has been discussed in detail in ref. \cite{Lu06}. It is worth to stress that the modified Chapman-Enskog expansion we are developing here differs formally from that used by Lutsko \cite{Lu06}. A short comparison of both expansion procedures is given in Appendix \ref{ap1}.

Next, the equation for the first order distribution, $f_{l}^{(1)}$, has to be considered. Collecting terms in the expansion of Eq.\ (\ref{2.6}) to first order in $\epsilon$, it is found
\begin{equation}
\label{3.18}
\partial_{t}^{(1)} f_{l}^{(0)}+ \partial_{t}^{(0)} f_{l}^{(1)}+ {\bm v} \cdot {\bm \nabla}^{(0)} f_{l}^{(1)}+ {\bm v} \cdot {\bm \nabla}^{(1)} f_{l}^{(0)} = \Lambda \left[ {\bm r}, {\bm v}|f_{eq} \right] f_{l}^{(1)}.
\end{equation}
The first order balance equation is
\begin{equation}
\label{3.19}
\partial_{t}^{(1)} n_{l}({\bm r},t) =-\left( {\bm \nabla}  \cdot {\bm J}_{l} \right)^{(1)} = - {\bm \nabla}^{(0)} \cdot  {\bm J}_{l}^{(1)},
\end{equation}
since it has been seen that ${\bm J}_{l}^{(0)}$ vanishes. Then, using that $f_{l}^{(1)}$ is a normal distribution, Eq. (\ref{3.18}) is equivalent to
\begin{equation}
\label{3.20}
\left( {\bm \nabla}^{(0)} \cdot {\bm J}_{l}^{(1)} \right) \varphi_{MB}({\bm v}) - v_{z} \frac{\partial n_{l,0}(z)}{\partial z} \nu ({\bm r},t) \frac{\partial f_{l}^{(1)}}{\partial n_{l}}
+ \Lambda \left[ {\bm r},{\bm v} |f_{eq} \right] f_{l}^{(1)} = \left[ {\bm v} \cdot {\bm \nabla} \nu({\bm r},t) \right] f_{l,eq}(z,{\bm v}).
\end{equation}
We look for solutions to this equation that must be proportional to ${\bm \nabla} \nu ({\bm r},t)$ by construction, i.e. they have the form
\begin{equation}
\label{3.21}
f_{l}^{(1)} ({\bm r}, {\bm v},t)= {\bm K}({\bm v}) \cdot {\bm \nabla} \nu ({\bm r},t),
\end{equation}
where ${\bm K}({\bm v})$ can also depend on ${\bm r}$ and $t$ through $n_{l}({\bm r},t))$, $n_{l,0}(z)$, and the derivatives of the latter. Use of this expression into the expansion of Eq. (\ref{2.21}) in powers of $\epsilon$ leads to
\begin{equation}
\label{3.22}
\boldsymbol{\mathcal{J}}_{l=}^{(1)} ({\bm r}_{=},t) = \int_{\sigma/2}^{h-\sigma/2} dz \int d{\bm v}\, {\bm v}_{=} {\bm K}({\bm v}) \cdot {\bm \nabla} \nu({\bm r},t).
\end{equation} 
Given the arbitrariness of ${\bm \nabla} \nu({\bm r},t)$, substitution of Eq.\ (\ref{3.21}) into Eq. (\ref{3.20}) yields
\begin{eqnarray}
\label{3.23}
\frac{\partial n_{l,0}}{\partial z} \varphi_{MB}({\bm v}) \nu ({\bm r},t) \frac{\partial}{\partial n_{l}({\bm r},t)} \int d {\bm v}^{\prime}\,  v_{z}^{\prime}  {\bm K}({\bm v}^{\prime})
&-& \frac{\partial n_{l,0}}{\partial z} \nu ({\bm r},t) v_{z} \frac{\partial {\bm K}({\bm v})}{\partial n_{l}({\bm r},t)}\,  \nonumber \\
&+& \Lambda \left[ {\bm r},{\bm v}|f_{eq} \right] {\bm K}({\bm v})
 = {\bm v} f_{l,eq} (z,{\bm v}).
\end{eqnarray}  
Integration of this equation over the velocity leads to a trivial identity, with no information about the function ${\bm K} ({\bm v})$. Now, it is noticed that the normal form of the distribution function requires that
\begin{equation}
\label{3.24}
K_{j}({\bm v}) = C_{j} v_{j} \Phi(v),
\end{equation}
(with no implicit summation over the repeated index) since ${\bm v}$ is the only vector of which ${\bm K}({\bm v})$ can depend. The coefficients $C_{j}$ may be functions of the density field of labeled particles, $n_{l}({\bm r},t)$ and also of $n_{l,0}(z)$ and its derivatives,  and $\Phi$ is an isotropic function of the velocity. Actually, because of dimensionality reasons and the definitions of ${\nabla ^{(0)}}$ and ${\bm \nabla}^{(1)}$, the quantities $C_{j}$ are expected to be proportional to $n_{l,0}(z)$. Substitution of Eq. (\ref{3.24}) into Eq. (\ref{3.23}), multiplication of the equation by $v_{i}$ and, finally, integration over ${\bm v}$ gives 
\begin{equation}
\label{3.25}
C_{j} \int d{\bm v}\, v_{i} \Lambda \left[ {\bm r},{\bm v}|f_{eq} \right] v_{j} \Phi (v)= \delta_{ij} \frac{k_{B}T}{m} n_{l,0}(z).
\end{equation}
By using the explicit form of the $\Lambda$ operator, given in Eq.\ (\ref{2.6}], it is verified that the left hand side of the above equation is diagonal, i.e. proportional to $\delta_{ij}$, as required by consistency.  In the next section, the diffusion equation for the density of labelled particles will be derived by solving Eq. (\ref{3.25}) in some approximation.

\section{The diffusion equation in the horizontal plane}
\label{s4}
A direct consequence of Eq. (\ref{3.24}) is that the horizontal flux of labeled particles given by Eq. (\ref{3.22}) takes the form
\begin{equation}
\label{4.1}
{\mathcal{J}}_{l=,i}^{(1)} ({\bm r}_{=},t) =  \int_{\sigma/2}^{h-\sigma/2} dz \int d{\bm v}\, C_{i} v_{i} ^{2} \Phi (v) \frac{\partial \nu ({\bm r},t)}{\partial r_{i}}\, ,
\end{equation}
where $i$ indicates any of the two components, $x$ or $y$, of a vector in the horizontal plane. Because of symmetry, it is $C_{x}=C_{y}= C_{=}$, so that the above equation is also equivalent to
\begin{equation}
\label{4.2}
\boldsymbol{\mathcal{J}}_{l=}^{(1)} ({\bm r}_{=},t) =  \frac{1}{2}  \int_{\sigma/2}^{h-\sigma/2} dz \int d{\bm v}\, C_{=} v_{=}^{2} \Phi(v) {\bm \nabla}_{=} \nu ({\bm r},t).
\end{equation}
To proceed, the first Sonine polynomial approximation for ${\bm K}({\bm v})$ will be considered \cite{RydL77}. In this  approximation, $\Phi (v) \sim \varphi_{MB}({\bm v})$, and 
\begin{equation}
\label{4.3}
\boldsymbol{\mathcal{J}}_{l=}^{(1)} ({\bm r}_{=},t) =  \frac{k_{B}T}{m} \int_{\sigma/2}^{h-\sigma/2} dz\, C_{=} {\bm \nabla_{=}} \nu ({\bm r},t).
\end{equation}
The determination of $C_{=}$ in the first Sonine approximation, requires to evaluate the integral (see Eq.\ (\ref{3.25}))
\begin{equation}
\label{4.4}
I_{=} \equiv \int d{\bm v}\, v_{i} \Lambda \left[ {\bm r},{\bm v}|f_{eq} \right] v_{i} \varphi_{MB} ({\bm v}).
\end{equation}
Again, it is $i=x,y$. By using the property given in Eq.\ (\ref{2.11}) and introducing the center of mass velocity, ${\bm G} \equiv ({\bm v}+{\bm v}_{1})/2$,  it is obtained
\begin{eqnarray}
\label{4.5}
I_{=}  & = & \frac{\sigma}{2} \int d{\bm g} \int_{\sigma/2}^{h-\sigma/2}  dz_{1} \int_{0}^{2\pi} d \psi\,  n_{0}(z_{1}) |{\bm g} \cdot \widehat{\bm \sigma} |^{2} \widehat{\sigma}_{x}g_{x}   \nonumber \\
&& \times  \left[ \Theta (-{\bm g}_{=} \cdot {\bm \sigma}_{=} ) - 2 \Theta ({\bm g} \cdot \widehat{\bm \sigma} ) \theta (-{\bm g}_{=} \cdot {\bm \sigma}_{=}) \right] \chi(g),
\end{eqnarray}
with
\begin{equation}
\label{4.6}
\chi(g) \equiv \left( \frac{m}{4 \pi k_{B}T} \right)^{3/2} e^{-\frac{mg^{2}}{4 k_{B} T}} \, .
\end{equation}
In the derivation of Eq.\ (\ref{4.5}) use has been made of the identity
\begin{equation}
\label{4.7}
\Theta \left( -\bm g \cdot \widehat{\bm \sigma} \right) = \Theta \left( - {\bm g}_{ =} \cdot {\bm \sigma}_{=}\right) - \Theta \left( {\bm g} \cdot \widehat{\bm \sigma} \right) \Theta \left( - {\bm g}_{ =} \cdot {\bm \sigma}_{=}\right) + \Theta \left( - {\bm g} \cdot \widehat{\bm \sigma} \right)  \Theta \left(  {\bm g}_{ =} \cdot {\bm \sigma}_{=}\right),
\end{equation}
as well as of the symmetry of the integrand under the change of ${\bm g}$ into $-{\bm g}$. The analytical evaluation of the integrals on the right hand side of Eq.\ (\ref{4.5}) seems quite involved, and to get a simple expression some kind of expansion has been considered. Details of the calculations are given in Appendix \ref{ap2}. The result is
\begin{equation}
\label{4.8}
I_{=} = -2\pi^{1/2} \sigma \left( \frac{k_{B}T}{m} \right)^{3/2}  \int_{\sigma/2}^{h-\sigma/2} dz_{1}\, n_{o}(z_{1}) (1- \cos^{2} \theta) + \mathcal{O} \left( B_{4} [z|n_{0}] \right),
\end{equation}
where
\begin{equation}
\label{4.9}
 B_{r} [z|n_{0}]  \equiv \int_{z/2}^{h-\sigma/2} dz_{1}\,  n_{0}(z_{1})\cos^{r} \theta .
\end{equation}
In the following, terms of order $B_{r}$ with $r \ge 4$ will be neglected. More about the meaning of this approximation will be said later on. Substitution of Eq.\ (\ref{4.8}) into Eq. (\ref{3.25}) gives
\begin{eqnarray}
\label{4.10}
C_{x} &=& -\left( \frac{k_{B}T}{m} \right)^{1/2} \frac{n_{l,0}(z)}{2 \pi^{1/2} \sigma} \left( \frac{N}{A} -B_{2}[z|n_{0}] \right)^{-1} \nonumber \\
& \approx & -\left( \frac{k_{B}T}{m} \right)^{1/2} \frac{n_{l,0}(z)A}{2 \pi^{1/2} \sigma N} \ \left( 1+ \frac{A}{N} B_{2}[z|n_{0}] \right).
\end{eqnarray}
Here it has been used that in the approximation we are considering, it is $A B_{2} \ll N$, as it will be explicitly shown below. When the above expression for $C_{x}$ is substituted into Eq. (\ref{4.2}), it is found that
\begin{equation}
\label{4.11}
\boldsymbol{\mathcal{J}}_{l=}^{(1)} ({\bm r}_{=},t)= - D_{0} \left[ {\bm \nabla}_{=} n_{l =} ({\bm r}_{=},t)+ \frac{A}{N} \int_{\sigma/2}^{h-\sigma/2} dz\, n_{l,0}(z) B_{2} [z|n_{0}] {\bm \nabla}_{=} \nu ({\bm r},t) \right].
\end{equation}
The coefficient
\begin{equation}
\label{4.12}
D_{0} = \frac{A}{2 N \pi^{1/2} \sigma}\, \left( \frac{k_{B} T}{m} \right)^{1/2}
\end{equation}
is the same as the equilibrium self-diffusion coefficient for a two-dimensional system of hard disks of diameter $\sigma$. with a superficial density $N/A$.

The structure of the expression for $C_{x}$ as given in Eq.\ (\ref{4.10}) deserves some comments. By construction, $f_{l}^{(1)}$ is assumed to have a normal form. Hence, $K_{x}$, introduced in Eq.\ (\ref{3.21}), and also $C_{x}$, defined in Eq.\ (\ref{3.24}), must also be normal. But the expression derived for $C_{x}$, and reported in Eq. (\ref{4.10}), depends explicitly on $z$ through $B_{2}$ and, consequently some apparent inconsistency shows up. Nevertheless, the analysis of the expression of $B_{2}$ given in Appendix \ref{ap3} shows that its $z$-dependence can be eliminated in favor of $n_{l,0}(z)$, so that  no contradiction exists.

The first term on the right hand side of Eq. (\ref{4.11}) has the form of the flux of particles describing a self-diffusion process in the horizontal plane.To express the other term in a similar way, it will be taken into account that the relative variation of $n_{0}(z)$ along its definition interval, $\sigma/2 < z < h-\sigma/2$, is small. This is illustrated, for instance, in Fig. 2 of ref. \cite{BMyG16}. Then, the following approximation is made
\begin{equation}
\label{4.13}
B_{2}[z|n_{0}] \rightarrow \overline{B}_{2} = \frac{1}{h-z} \int_{\sigma/2}^{h-\sigma/2} dz \int_{\sigma/2}^{h-\sigma/2} dz_{1} \frac{N(z-z_{1})^{2}}{A(h-\sigma)\sigma^{2}}
= \frac{N (h-\sigma)^{2}}{6A \sigma^{2}}\,.
\end{equation}
Then, this estimate consists in replacing the equilibrium  density of the fluid by its average along the vertical direction and, afterwards, in doing the same with the resulting function of $z$. Substitution of Eq.\ (\ref{4.13})  into Eq.\ (\ref{4.11}) gives
\begin{equation}
\label{4.14}
 \boldsymbol{\mathcal{J}}_{l=}^{(1)} ({\bm r}_{=},t)= -D {\bm \nabla}_{=} n_{l =} ({\bm r}_{=},t),
 \end{equation}
 with the modified diffusion coefficient given by
 \begin{equation}
 \label{4.15}
 D=D_{0} D^{*} (h),
 \end{equation}
 \begin{equation}
 \label{4.16}
 D^{*}(h) = 1+\frac{1}{6} \left( \frac{h}{\sigma}-1 \right)^{2}\, .
 \end{equation}
 Combination of Eqs.\ (\ref{2.19}) and (\ref{4.14}) results in the diffusion equation of labeled particles ias projected on the horizontal plane,
 \begin{equation}
 \label{4.17}
 \frac{\partial}{\partial t}\, n_{l=}({\bm r}_{=},t) = D \nabla^{2}_{=}  n_{l=}({\bm r}_{=},t).
 \end{equation}
 Therefore, the effective self-diffusion coefficient associated to the quasi-two-dimensional motion observed when the system is seen from above, increases as the separation $h$  between the two plates increases. This result has been derived when $h$ is close to $\sigma$, and expected to be qualitatively valid up to $h=2 \sigma$. A point to be noticed is that the correction to the two-dimensional bulk diffusion coefficient in Eq. (\ref{4.16}) is independent from the density and, therefore, it can not be interpreted as a higher order in the density effect.

\section{Molecular dynamics simulation results}
\label{s5}
In order to check the accuracy of the theoretical predictions obtained in the previous sections, molecular dynamics (MD) simulations of a system of hard spheres have been performed. The simulation technique was based on the ``event driven'' algorithm \cite{AyT87}. The domain was a square base rectangular parallelepiped  limited by the two plates, and doubly periodic boundary conditions in the plane parallel to the plates were employed. Initially, all the particles were uniformly distributed and the velocity distribution was a Gaussian with temperature $T(0)$.  It was checked that the system remained homogeneous when projected on the horizontal plane and  that, after a transient, a stationary equilibrium state was reached.  The results to be reported in the following correspond to a system of $N=500$ particles and the value of $A$ was such that $N/A = 0.019 \sigma^{2}$. 

\begin{figure}
\includegraphics[scale=0.5,angle=0]{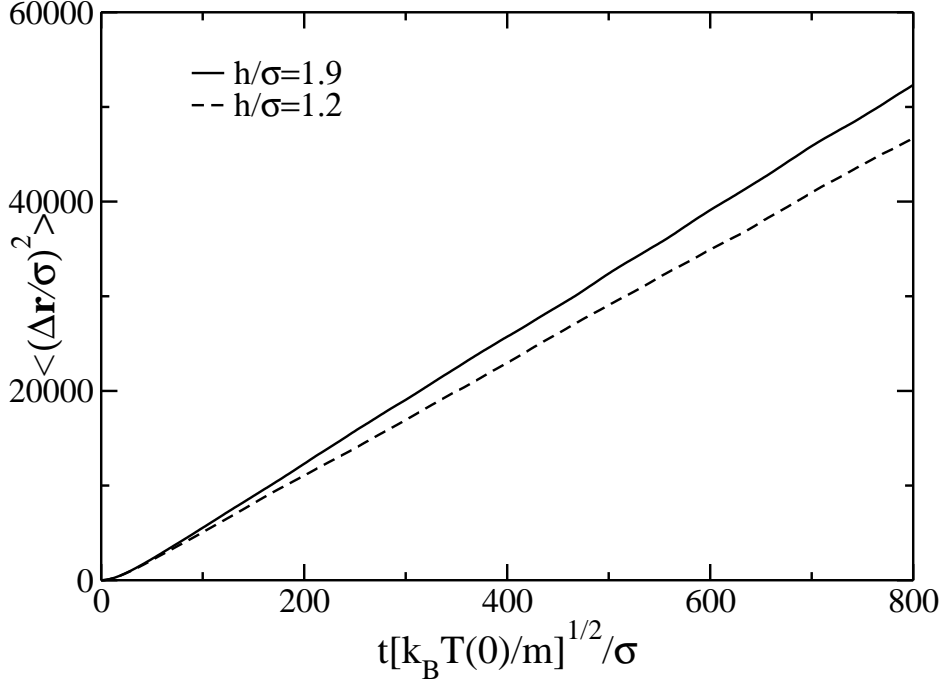}
\caption{Mean square displacement of the particles as a function of time for two values of the separation of the two plates, as indicated in the inset. Both quantities are measured in the dimensionless units indicated in the labels.}
\label{fig1}
\end{figure}

\begin{figure}
\includegraphics[scale=0.5,angle=0]{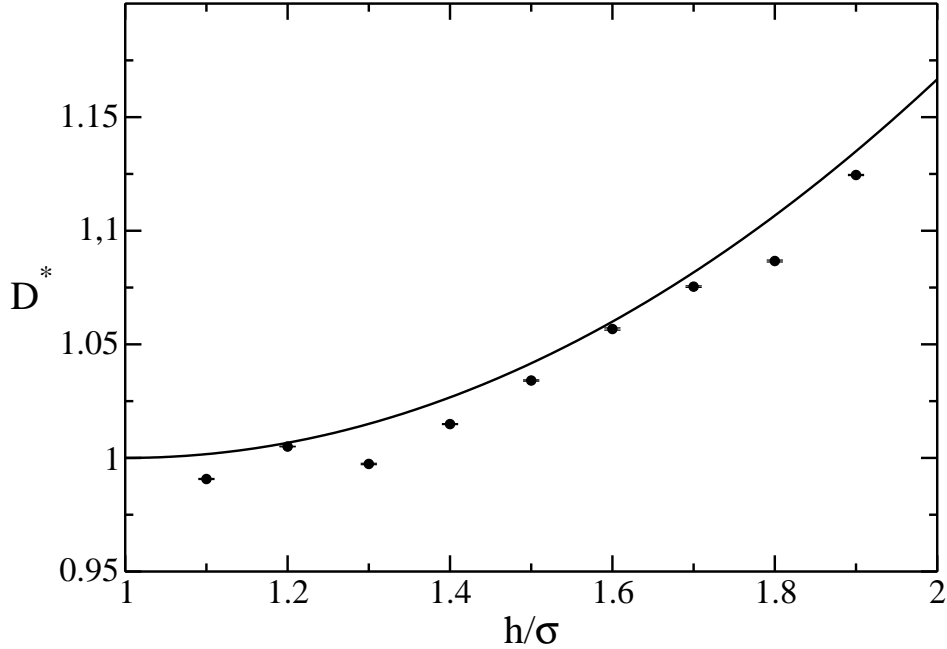}
\caption{The dimensionless reduced self-diffusion coefficient $D^{*}$, defined in Eq. (\ref{4.15}), as a function of the  distance $h$ between the two confining plates. The latter is measured in units of the diameter of the particles $\sigma$. The solid line is the theoretical prediction given by Eq. (\ref{4.16}), while the symbols have been obtained by MD simulations of a system of hard spheres, as described in the main text.}
\label{fig2}
\end{figure}

The diffusion equation (\ref{4.17}) implies that the mean square deviations of the projected position ${\bm r}_{=}$ of labelled particles after a time interval $t$ is
\begin{equation}
\label{5.1}
\langle \left( \Delta {\bm r}_{=} \right)^{2} ;t \rangle = 4 D t.
\end{equation}
The angular brackets denote average over trajectories of different labelled particles. The method used to measure the diffusion coefficient ion the simulation is directly based on the above equation. In Fig. \ref{fig1}, $\langle \left( \Delta {\bm r}_{=} \right)^{2} \rangle$ is plotted as a function of time. The data  have been obtained by averaging over all the particles and also over 20 different trajectories of the system, in order to reduce the statistical uncertainties. Results for two different  separation $h$ of the two plates are reported.  They correspond to values close to the lowest and highest limits for which the theory is expected to  apply. It is observed that, after a short transient period, of the order of a few collisions per particle, the mean square displacement becomes a linear function of time, indicating the diffusive nature of the motion.  Similar results have been obtained for other  values of $h$.  From each value of the slope in the linear region, the diffusion coefficient $D$ has been computed by means of Eq. (\ref{5.1}). The comparison with the theoretical prediction, Eqs.  (\ref{4.15}) and (\ref{4.16}), is presented in Fig. \ref{fig2}. 
 The solid line is the theoretical prediction and the symbols correspond to values obtained from the simulation data.  The error bars are  of the same size as the symbols employed to represent the data. A good agreement is observed for all the range of values of the distance $h$. Actually, it is surprising that no systematic increase of the discrepancy between theory and simulation shows up as $h$ approach the limiting value $2 \sigma$, since in the theoretical analysis an expansion in $(h-\sigma)^{2}/\sigma^{2}$ was carried out, and only the leading term was kept (see Eq.\ (\ref{4.8}), and the discussion following it). The value of the diffusion coefficient changes by an amount of the order of $15 \%$ in the range of $h$ considered. To properly value the reported results, it is emphasized that when varying the separation $h$,  the effective two-dimensional number density $N/A$ is kept constant and, consistently, the three-dimensional  density is different for each value of $h$.

\section{Discussion}
\label{s6}
In this paper, the self-diffusion process in a Q2D system of hard spheres has been analyzed starting from a Boltzmann-Lorentz kinetic equation. The system is confined by means of two hard parallel infinite plates, separated a distance smaller that twice the diameter of a particle. It has been shown that the standard Chapman-Enskog procedure to derive a normal solution of a kinetic equation can be extended in a quite natural way to systems in which the equilibrium state exhibits density gradients. A particular simplifying feature of dealing with the equilibrium state is that the expansion can be carried out in such a way that the zeroth order distribution in the expansion is obtained from the equilibrium one by replacing the equilibrium density by the actual density profile. This does not happens when transport around a general non-equilibrium state is considered \cite{Lu06}. To study the quality of the theory developed and the approximations made in the calculations, comparison with molecular dynamics simulations have been presented. The simulation data confirm the presence of diffusive behavior in the horizontal plane, i.e. the mean square displacement growths linearly in time for large times, and also the accuracy of the expression for the self-diffusion coefficient. 

To put the present work in a proper context, it is worth to emphasize that no diffusion process is expected to happen in the vertical direction. In this sense, it differs from those investigations in which the possibility of projecting three-dimensional diffusion processes on one direction is investigated. This is the case, for instance, when considering diffusion in a channel of varying cross section and it is described as a one-dimensional diffusion past an entropy barrier determined by the channel width  \cite{Zw92,RyR01}. On the other hand, it is clear that the effect of  strong confinement can be associated to the presence of an entropic force that restricts the three-dimensional motion to a Q2D one. Therefore, a natural extension of the present work is to study a strong confinement in which the separation between the two parallel hard surfaces varies. Of course, this requires to modify the starting kinetic equation, but the way to do it seems quite clear.

\acknowledgments

This research was supported by the Ministerio de Econom\'{\i}a, Industria  y Competitividad  (Spain) through Grant No. FIS2017-87117-P (partially financed by FEDER funds).

\appendix

\section{Comparisson of the expansion carried out in this paper and that used by Lutsko \cite{Lu06}}
\label{ap1}

Lutsko's expansion adapted to the present problem of self-diffusion would be as follows. First, define $\delta n_{l}({\bm r},t)$ by
\begin{equation}
\label{ap1.1}
n_{l}({\bm r},t)= n_{l,0} (z) + \delta n_{l}({\bm r},t).
\end{equation}
Next the gradient operator is decomposed as
\begin{equation}
\label{ap1.2}
{\bm \nabla} n_{l} ({\bm r},t) \equiv {\bm \nabla}^{(0) \prime} n_{l}({\bm r},t) + \epsilon {\bm \nabla}^{(1)\prime} n_{l}({\bm r},t),
\end{equation}
with
\begin{equation}
\label{ap1.3}
{\bm \nabla}^{(0) \prime} n_{l}({\bm r},t) \equiv \frac{\partial n_{l,0}(z)}{\partial z}\ \widehat{\bm e}_{z},
\end{equation}
\begin{equation}
\label{ap1.4}
{\bm \nabla}^{(1) \prime} n_{l}({\bm r},t) \equiv {\bm \nabla} \delta n_{l}({\bm r},t).
\end{equation}
A prime symbol is used to differentiate these definitions from  Eqs. (\ref{3.2})-(\ref{3.4}), that define the modified Chapmn-Enskog expansion as formulated here.
Both expansions can be easily related by noting that
\begin{equation}
\label{ap1.5}
\nu({\bm r},t) = 1+ \frac{\delta n_{l}({\bm r},t)}{n_{l,0}(z)}
\end{equation}
and, therefore,
\begin{equation}
\label{ap1.6}
{\bm \nabla}^{(1)} n_{l}({\bm r},t) =  {\bm \nabla} ^{(1) \prime} n_{l}({\bm r},t)- \frac{\delta n_{l}({\bm r},t) {\bm \nabla}^{(0) \prime} n_{l}({\bm r},t)}{n_{l,0}(z)}\, .
\end{equation}
Of course, the relation between the operators ${\bm \nabla}^{(0)}$ and ${\bm \nabla}^{(0) \prime}$ is the same, but changing the minus sign by the plus sign on the right hand side in Eq. (\ref{ap1.6}). In principle, no physical or mathematical reason seems to exist to prefer any of the two expansions, being, therefore, just a matter or convenience for the specific problem at hand. For the case of self-diffusion in a system having and inhomogeneous equilibrium state, we have found more convenient the expansion based on the decomposition given in Eqs. (\ref{3.2})-(\ref{3.4}), because it leads to a quite simple identification of the zeroth order Chapman-Enskog solution. Moreover, as mentioned in the main text, this choice is closely related with the concept of local equilibrium  for systems having a non-uniform equilibrium state.

\section{Evaluation of the integral defined in Eq. (\ref{4.5})}
\label{ap2}
To begin with, the expansion
\begin{equation}
\label{ap2.1}
|{\bm g} \cdot \widehat{\bm \sigma}|^{2} = |{\bm g}_{=} \cdot {\bm \sigma}_{=} |^{2} + |g_{z} \widehat{\sigma}_{z} |^{2} + 2 {\bm g}_{=} \cdot{\bm \sigma}_{=} g_{z} \widehat{\sigma}_{z}
\end{equation}
is substituted on the right hand side of Eq. (\ref{4.5}) to get
\begin{eqnarray}
\label{ap2.2}
I_{=}  & = & \frac{\sigma}{2} \int d{\bm g} \int_{\sigma/2}^{h-\sigma/2}  dz_{1} \int_{0}^{2\pi} d \psi\,  n_{0}(z_{1}) \widehat{\sigma}_{x}g_{x}  \left[ |{\bm g}_{=} \cdot {\bm \sigma}_{=} |^{2} + |g_{z} \widehat{\sigma}_{z} |^{2} + 2 {\bm g}_{=} \cdot{\bm \sigma}_{=} g_{z} \widehat{\sigma}_{z} \right] \nonumber \\
&& \times  \left[ \Theta (-{\bm g}_{=} \cdot {\bm \sigma}_{=} ) - 2 \Theta ({\bm g} \cdot \widehat{\bm \sigma} ) \theta (-{\bm g}_{=} \cdot {\bm \sigma}_{=}) \right] \chi(g),
\end{eqnarray}
The evaluation of this expression requires to consider six integrals. Let us study each of them separately. The first one is
\begin{eqnarray}
\label{ap2.3}
I_{=}^{(1)}  &\equiv&  \frac{\sigma}{2} \int d{\bm g} \int_{\sigma/2}^{h-\sigma/2}  dz_{1} \int_{0}^{2\pi} d \psi\,  n_{0}(z_{1}) \widehat{\sigma}_{x}g_{x} |{\bm g}_{=} \cdot {\bm \sigma}_{=} |^{2} \Theta (-{\bm g}_{=} \cdot {\bm \sigma}_{=} ) \chi(g) \nonumber \\
&&=  \frac{\sigma}{4} \int d{\bm g} \int_{\sigma/2}^{h-\sigma/2}  dz_{1} \int_{0}^{2\pi} d \psi\,  n_{0}(z_{1}) ({\bm g}_{=} \cdot {\bm \sigma}_{=})^{3} \Theta (-{\bm g}_{=} \cdot {\bm \sigma}_{=} ) \chi(g).
\end{eqnarray}
In the last transformation, we have interchanged $v_{x}$ and $v_{y}$ and changed $\psi$ into $\pi/2-\psi$. The latter is equivalent to interchange $\widehat{\sigma}_{x}$ and $\widehat{\sigma}_{y}$. Next, we use that ${\bm \sigma}_{=}= (\sin \theta) \widehat{\bm \sigma}_{=}$, where $\widehat{\bm \sigma}_{=}$ is a unit vector in the horizontal plane. Moreover, since $0<\theta<\pi$, it is $\Theta(-g \cdot {\bm \sigma}_{=}) = \Theta (-{\bm g} \cdot \widehat{\bm \sigma}_{=})$. It follows that Eq.\ (\ref{ap2.3}) is equivalent to
\begin{equation}
\label{ap2.4}
I_{=}^{(1)}= \frac{\sigma}{4} \int d{\bm g} \int_{\sigma/2}^{h-\sigma/2}  dz_{1} \int_{0}^{2\pi} d \psi\,  n_{0}(z_{1})  (\sin \theta)^{3} ({\bm g}_{=} \cdot \widehat{\bm \sigma}_{=})^{3} \Theta (-{\bm g}_{=} \cdot \widehat{\bm \sigma}_{=} ) \chi(g).
\end{equation}
Carrying out the angular and velocity integrals, it is found
\begin{equation}
\label{ap2.5}
I_{=}^{(1)}=-2\pi^{1/2} \sigma \left( \frac{k_{B}T}{m} \right)^{3/2} \int_{\sigma/2}^{h-\sigma/2} dz_{1}\, n_{0}(z_{1}) \sin^{3} \theta.
\end{equation}
The next contribution to $I_{=}$ in Eq. (\ref{ap2.2}) to be considered is
\begin{equation}
\label{ap2.6}
I_{=}^{(2)} \equiv \frac{\sigma}{2} \int d{\bm g} \int_{\sigma/2}^{h-\sigma/2}  dz_{1} \int_{0}^{2\pi} d \psi\,  n_{0}(z_{1}) \widehat{\sigma}_{x}g_{x}  \widehat{\sigma}_{z}^{2}  g_{z}^{2} \Theta (-{\bm g}_{=} \cdot {\bm \sigma}_{=} ) \chi(g).
\end{equation}
By using the same method as for the previous integral, it is obtained
\begin{equation}
\label{ap2.7}
I_{=}^{(2)} = -\sigma \pi^{1/2} \left( \frac{k_{B}T}{m} \right)^{3/2} \int_{\sigma/2}^{h-\sigma/2} dz_{1}\, n_{0}(z_{1}) \sin \theta \cos^{2} \theta.
\end{equation}
The third integral to evaluate is
\begin{equation}
\label{ap2.8}
I_{=}^{(3)} \equiv \sigma \int d{\bm g} \int_{\sigma/2}^{h-\sigma/2}  dz_{1} \int_{0}^{2\pi} d \psi\,  n_{0}(z_{1}) \widehat{\sigma}_{x}g_{x}  ({\bm g}_{=} \cdot {\bm \sigma}_{=}) g_{z} \widehat{\sigma}_{z}\Theta (-{\bm g}_{=} \cdot {\bm \sigma}_{=} ) \chi(g) ,
\end{equation}
and it vanishes since the integrand is an odd function of $g_{z}$. To analyze the next term,
\begin{equation}
\label{ap2.9}
I_{=}^{(4)} \equiv - \sigma  \int d{\bm g} \int_{\sigma/2}^{h-\sigma/2}  dz_{1} \int_{0}^{2\pi} d \psi\,  n_{0}(z_{1}) \widehat{\sigma}_{x}g_{x}  ({\bm g}_{=} \cdot {\bm \sigma}_{=})^{2} \Theta ({\bm g} \cdot \widehat{\bm \sigma} ) \Theta (-{\bm g}_{=} \cdot {\bm \sigma}_{=} ) \chi(g), 
\end{equation}
the formal expansion
\begin{equation}
\label{ap2.10}
\Theta ({\bm g} \cdot \widehat{\bm \sigma}) = \Theta ({\bm g}_{=} \cdot {\bm \sigma}_{=}) + \delta ({\bm g}_{=} \cdot {\bm \sigma}_{=}) g_{z} \widehat{\sigma}_{z} + \frac{1}{2} \delta^{\prime} ({\bm g}_{=} \cdot {\bm \sigma}_{=}) g_{z}^{2} \widehat{\sigma}_{z}^{2} + \ldots,
\end{equation}
is used. Here $\delta^{\prime}(x)$ denotes  the derivative of the delta function. When the expansion is introduced in Eq.\ (\ref{ap2.9}) it is seen that the first non-vanishing contribution contains $\widehat{\sigma}_{z}^{4}$, i.e.
\begin{equation}
\label{ap2.11}
I_{=}^{(4)} = \mathcal{O} \left[\int_{z/2}^{h-\sigma/2} dz_{1}\,  n_{0}(z_{1}) \cos^{4} \theta \right]
\end{equation}
By using similar arguments, he same conclusion is reached for the remaining two contributions to $I_{=}$. Finally, to get a consistent approximation,  the expansion 
\begin{equation}
\label{ap2.12}
\sin^{3} \theta = 1- \frac{3}{2} \cos^{2} \theta + \mathcal{O}( \cos^{4} \theta)
\end{equation}
is employed in Eq. (\ref{ap2.5}), while in Eq.\ (\ref{ap2.7}) one uses
\begin{equation}
\label{ap2.13}
\sin \theta =1+ \mathcal{O} ( \cos^{2} \theta).
\end{equation}
Then, Eq.\ (\ref{4.8}) follows by collecting the six contributions to the integral as derived above.

\section{The `normal'' property of $B_{2} [z|n_{0}]$.}
\label{ap3}
The global equilibrium density given by Eq.\ (\ref{2.3}) verifies the equation
\begin{equation}
\label{ap3.1}
\frac{\partial}{\partial z}\, \ln n_{0}(z) = 2 \pi \int_{\sigma/2}^{h-\sigma/2} dz_{1} n_{0}(z_{1}) (z-z_{1}),
\end{equation}
as it can be checked by direct substitution. Then, it follows that
\begin{equation}
\label{ap3.2}
\frac{\partial}{\partial z} B_{2} [z|n_{0}] = \frac{2}{\sigma^{2}} \int_{\sigma/2}^{h-\sigma/2} dz_{1} n_{0}(z_{1}) (z-z_{1}) = \frac{1}{\pi \sigma^{2}} \frac{\partial }{\partial z} \ln n_{0} (z),
\end{equation}
and integration of this equation with respect to $z$ gives
\begin{equation}
\label{ap3.3}
B_{2}[z|n_{0}]= (\pi \sigma^{2})^{-1} \ln n_{0}(z) + D.
\end{equation}
The integration constant $D$ can be determined, for instance, by particularizing the expression doe $z=h/2$. A simple calculation shows that
\begin{equation}
\label{ap3.4}
B_{2}[h/2|n_{0}]= - \frac{N}{2 Aa \sigma^{2}} + \frac{h}{2 a \sigma} n_{0} \left( \frac{h}{2} \right).
\end{equation}
Use of this result  into Eq. (\ref{ap3.3}) leads to the identification
\begin{equation}
\label{ap3.5}
C= - \frac{1}{2\pi \sigma^{2}} + \frac{h-\sigma}{2a \sigma^{2}} n_{0} \left( \frac{h}{2} \right) -\frac{1}{\pi \sigma^{2}} \ln n_{0} \left( \frac{h}{2} \right).
\end{equation}
Finally, substitution of  this values in Eq. (\ref{ap3.3}) provides the expression of $B_{2}$ in terms of the equilibrium density,
\begin{equation}
\label{ap3.6}
B_{2}[z|n_{0}] = \frac{1}{\pi \sigma^{2}} \ln \frac{n_{0}(z)}{n_{0}(h/2)} -\frac{1}{2 \pi \sigma^{2}} \left( 1- \frac{n_{0} (h/2)}{\overline{n}_{0}} \right),
\end{equation}
where $\overline{n}_{0} \equiv N/A(h-\sigma)$. This proofs that the expression derived in the main text for $f_{l}^{(1)}$ is consistent with the formulated modified Chapman-Enskog expansion, in the sense that it is a normal solution, with the meaning of normal distribution as formulated in this paper.

\end{document}